# Towards Collective Superintelligence: Amplifying Group IQ using Conversational Swarms

Louis Rosenberg
Unanimous AI
Pismo Beach, California
Louis@Unanimous.ai

Gregg Willcox
Unanimous AI
Seattle, Washington
Gregg@Unanimous.ai

Hans Schumann
Unanimous AI
San Francisco, California
Hans@Unanimous.ai

Ganesh Mani
Carnegie Mellon University
Pittsburgh, Pennsylvania
ganeshm@andrew.cmu.edu

*Abstract*— Swarm Intelligence (SI) is a natural phenomenon that enables biological groups to amplify their combined intellect by forming real-time systems. Artificial Swarm Intelligence (or Swarm AI) is a technology that enables networked human groups to amplify their combined intelligence by forming similar systems. In the past, swarm-based methods were constrained to narrowly defined tasks like probabilistic forecasting and multiple-choice decision making. A new technology called Conversational Swarm Intelligence (CSI) was developed in 2023 that amplifies the decision-making accuracy of networked human groups through natural conversational deliberations. The current study evaluated the ability of real-time groups using a CSI platform to take a common IQ test known as Raven's Advanced Progressive Matrices (RAPM). First, a baseline group of participants took the Raven's IQ test by traditional survey. This group averaged 45.6% correct. Then, groups of approximately 35 individuals answered IQ test questions together using a CSI platform called Thinkscape. These groups averaged 80.5% correct. This places the CSI groups in the 97$^{th}$ percentile of IQ test-takers and corresponds to an effective IQ increase of 28 points (p<0.001). This is an encouraging result and suggests that CSI is a powerful method for enabling conversational collective intelligence in large, networked groups. In addition, because CSI is scalable across groups of potentially any size, this technology may provide a viable pathway to building a Collective Superintelligence.

## I. Introduction

Many natural species have independently evolved the ability to amplify their collective intelligence by forming real-time systems such as bird flocks, bee swarms, and fish schools. This is commonly called Swarm Intelligence (SI) and it enables many social organisms to make group decisions that are significantly smarter than the individuals could achieve on their own [1]. In 2015, a technology called Artificial Swarm Intelligence (or Swarm AI) was developed to enable networked human groups to make decisions by forming real-time systems modeled on the dynamics of biological swarms [2]. These Swarm AI systems have been shown to significantly amplify the accuracy of groups decisions across a variety of common tasks, from forecasting financial markets and sporting events, to predicting sales, inventory, and consumer insights. [3 - 7].

While traditional Swarm AI technology has proven effective for many applications, the use-cases have been limited because questions had to be formatted as numerical estimates, such as probabilistic forecasts, or multiple-choice selections among sets of predefined options. To address these limitations, researchers developed a new method in 2023 called Conversational Swarm Intelligence (CSI) that combines the principles of Swarm AI with the power of Large Language Models (LLMs) [8,9].

The goal of CSI technology is to empower large, networked groups of potentially any size to hold real-time conversational deliberations that are thoughtful, productive, and amplify the group's collective intelligence on open-ended problems. This is a challenging goal because real-time conversations are optimally efficient in small groups of only 4 to 7 individuals and rapidly lose effectiveness with increasing size [10]. To solve this, CSI takes its inspiration from the behavior of fish schools [11]. That's because large schools of fish can make rapid decisions in life-or-death situations without a central authority mediating the process. Evolution achieved this by enabling each individual to hold a "conversation" with a small subset of nearby fish using a unique organ called a *lateral line* that detects faint pressure changes as neighbors adjust their direction and speed. And because each local subset overlaps other subsets, information quickly propagates within the full population. This enables the emergent property of Swarm Intelligence and allows thousands of individuals to quickly converge on unified decisions that are critical for survival [12, 13].

CSI emulates the communication structure of a fish school by breaking large human groups into a network of overlapping subgroups, each sized with 4 to 7 members for optimal real-time conversational deliberation. The problem, of course, is that humans did not evolve with the ability to hold conversations in overlapping subgroups. After all, if we had that ability – any cocktail party would become a swarm intelligence with information propagation around the room. This does not happen because humans evolved the opposite ability – to focus *only* on our local group and tune out conversational distractions from neighboring groups. This is called the "cocktail party effect" and it keeps us focused on local deliberations. [18]

To overcome this barrier in human abilities, CSI technology uses artificial agents powered by Large Language Models (LLMs) to enable the real-time overlap among deliberating groups [8, 9, 11, 17]. Specifically, CSI works by breaking a large group into a network of subgroups such that an LLM-powered conversational agent is inserted into each of the subgroups and tasked with observing the deliberation in that group, distilling the salient content, and passing critical points to other subgroups where its local AI agent will express the points as a natural part of the conversation. Of course, this process of observing, passing, and expressing happens in all rooms simultaneously, enabling conversational content to smoothly propagate. Using this novel CSI architecture, 25, 250 or even 2,500 people can hold a real-time deliberation, sharing views and ideas, debating options and alternatives, and converging in unison on solutions that garner maximal support.

An example CSI structure is shown in Figure 1. It represents a group of 98 real-time participants divided into a network of 14 subgroups, each one populated with 7 human users and one artificial agent. While the image implies that each subgroup can pass information to two other subgroups in the network, the actual model used was *fully connected*, meaning that the AI agent in each subgroup could potentially pass content to any other subgroup in the network depending on *a matchmaking subsystem* that considers the conversational dynamics in each available subgroup at that time. Because this structure is highly scalable, it could be used to connect thousands or even millions of users in real-time, either using a flat network structure as shown, or a nested network structure. Either way, the scalability means it could provide a pathway to collective superintelligence.

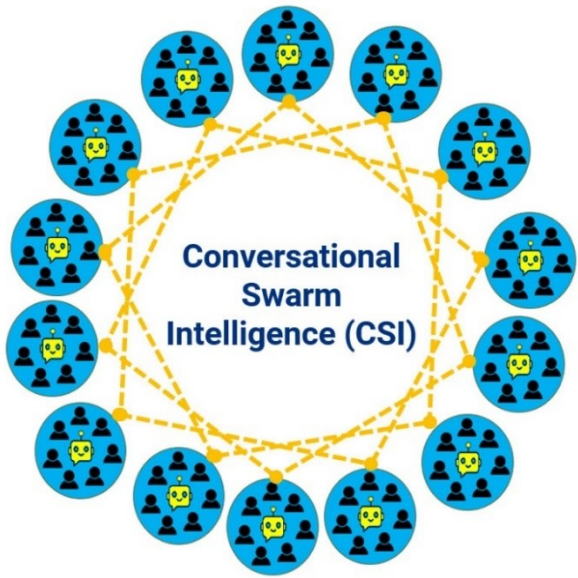

**Fig. 1.** Architecture for a Conversational Swarm Intelligence with AI agents assigned to each subgroup for passing and receiving conversastional content.

By facilitating large groups to discuss complex problems in real-time, the CSI structure enables participants with a wide range of knowledge, wisdom, and insights to consider broad, open-ended problems, and debate a wide array of solutions that organically emerge. In general, strongly supported ideas propagate faster through the network than weakly supported ideas. And yet, because the process is deliberative, with real-time reactions to every comment made, arguments accumulate in favor or against each assertion, enabling weakly supported ideas to overcome early skepticism, if warranted, while initially favored ideas can fade over time as they are vetted. And because every assertion is databased in real-time by the CSI system, documenting the arguments made in support and opposition, the system can generate detailed forensic reports that reveal how and why each decision was reached.

In this way, CSI not only promotes convergence on strong solutions, it captures the reasons and rationales that underlie the process. In addition, CSI is designed to reduce the impact of social influence bias because each member is only directly exposed to comments by a small number of others in real-time, reducing the impact of early views and/or strong personalities on the full population. In this way, CSI combines the intelligence amplification benefits of large groups with the deliberative reasoning of small groups.

Although a newly developed technology, a number of published studies already suggest that CSI is a powerful method for enhancing collaboration, communication, and collective intelligence among networked groups. In one early study at Carnegie Mellon in 2023, real-time groups of 25 participants were tested using the Thinkscape CSI platform and compared to standard centralized chat. The participants in the CSI structure produced 30% more contributions ($p<0.05$) than those using standard chat and 7.2% less variance, indicating that users participated more evenly when using CSI [8].

In a larger study, groups of 48 users were tasked with brainstorming and debating a topic rooted in current events – *the impact of AI on jobs.* The participants using CSI contributed 51% more content ($p<0.001$) compared to those using standard centralized chat. In addition, CSI showed 37% less difference in contribution between the most vocal and least vocal users, indicating that CSI fosters more balanced deliberations. In addition, a large majority of participants preferred the CSI platform over standard chat ($p<0.05$) and reported feeling more impactful when using the Thinkscape system ($p<0.01$) [9].

In another study, a real-time deliberative group of 80 participants was tested in the Thinkscape platform to assess the ability of CSI to generate qualitative insights regarding a set of political candidates running for office in the United States in 2024. After a short period of chat-based deliberation, the group converged on a preferred candidate and surfaced over 200 reasons for supporting that candidate. The maximally supported solution converged globally, garnering a statistically significant sentiment level within only six minutes ($p<0.001$) [11,12].

In the largest study to date, 245 users engaged in a single largescale text-chat conversation using the Thinkscape platform. The group was tasked with estimating the number of gumballs in a jar by viewing a photograph online. The CSI method partitioned the 245 participants into 47 subgroups of 5 or 6 members while AI agents passed conversational content around the network [16]. The estimates generated using Thinkscape were compared to a traditional survey-based aggregation across the same population of users. In addition, GPT-4.0 was given the same photo and tasked with estimating the gumballs. The group using CSI outperformed the average individual, the traditional wisdom of crowd, and GPT-4.0. In fact, the CSI estimate had a 50% smaller error than the survey based WoC technique, a surprising result [17].

While prior studies have clearly shown that groups can increase their collective intelligence using CSI, no prior study has tested the amplification of intelligence using standardized IQ test. The objective of the new study described below is to explore if groups can amplify their IQ when conversationally deliberating in connected subgroups mediated by CSI.

## II. STUDY OF IQ AMPLIFICATION

To assess if networked human groups can hold real-time deliberative conversations using a CSI networking structure and to quantify the degree to which the technology can amplify the group's collective intelligence, sets of approximately 35 people (randomly sourced using a commercial sample provider) were paid a small fee to login to the Thinkscape platform. Each group was tasked with answering standard IQ test questions through real-time collaborative deliberation. The Thinkscape platform automatically divided the 35-person groups into 7 subgroups of 5 people. Each subgroup was assigned an AI agent, as described above, to observe insights generated by that subgroup and share those insights with other AI agents within other subgroups. Those other agents express those insights conversationally within those local deliberations while also observing and sharing insights with other subgroups. This creates an overlapping conversational structure, turning the 7 local conversations into a unified global conversation that can converge on solutions that maximized support and amplify collective intelligence.

For clarity, it's important to note that using the CSI structure, each individual participant was only able to converse with the other 4 members of their subgroup and with the assigned AI agent. The AI agents <u>did not</u> introduce any content into the system – they only passed and received conversational insights from other subgroups, enabling the full 35-person group to function as a unified conversational system. In addition, a baseline group of 35 people were tasked with taking the IQ test as isolated individuals using a standard survey. Participants were disqualified for randomly guessing or cheating based on their pattern of survey responses and the elapsed time of their effort.

In this study, the research team used IQ test questions sourced from a popular intelligence test known as the Raven's Advanced Progressive Matrices (RAPM). This instrument measures the deductive reasoning ability in test-takers. The RAPM test was chosen because of its acceptance as a reputable measure of IQ and because of its simple visual format – all questions are presented as a set of images with a missing image that completes a presented pattern. In addition, prior studies have shown the RAPM test gives consistent results when administered to paid participants [14]. An example question from the RAPM test is shown below in Figure 2 [15].

Participants were given up to 4 minutes to answer each question. This means that each 35-person group had only 4 minutes to hold a networked real-time deliberation across subgroups and converge on an answer using Thinkscape.

## III. DATA AND ANALYSIS

The individual IQ test surveys (filtered for bad actors) were assessed to provide a baseline for paid participants sourced from a commercial sample provider. The average survey participant scored approximately half the questions correct (45.7%) and were assigned a nominal IQ score of 100. The participant groups used for the Thinkscape (CSI) trials were randomly sourced from the same provider and can be assumed to also have a distribution with an average IQ of approximately 100.

When using Thinkscape, the CSI group debated each IQ test question using text-based chat in their local subgroups, while AI agents passed content across the set of 7 subgroups. That content <u>only</u> reflected views surfaced within subgroups and introduced no other information. Real-time natural language processing (NLP) built into Thinkscape assessed the strength of conviction for each of the eight possible choices in each question, allowing the system to monitor in real-time which answer options were preferred by the full population. At the end of the allotted time, the answer with the greatest conversational sentiment was selected as the groupwise answer and scored accordingly.

After all sessions were scored, the "effective IQ" of the average Thinkscape group was calculated as a function of the average accuracy and standard deviation on the test. According to the standard IQ formula, $\mu$ is the mean individual score on the test, $\sigma$ is the standard deviation of individual scores on the test, and X is the score to convert to an IQ as follows:

$$IQ(X) = 100 + 15 * \frac{(X - \mu)}{\sigma} \quad [Eq. 1]$$

## IV. RESULTS

Looking first at the baseline surveys, the average individual test-taker scored 45.7% correct. The distribution of individuals is shown in Figure 3 (orange bars) inside of a normal curve fit with the same mean (45.7%) and standard deviation (18.6%) as the sample distribution of individuals for reference. This curve is used for the basis of future IQ calculations.

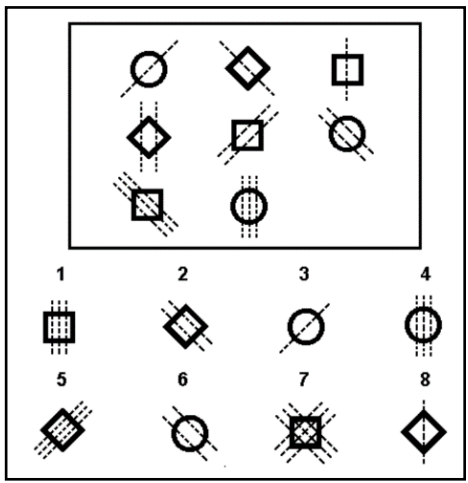

**Fig. 2.** Sample Question from RAPM test

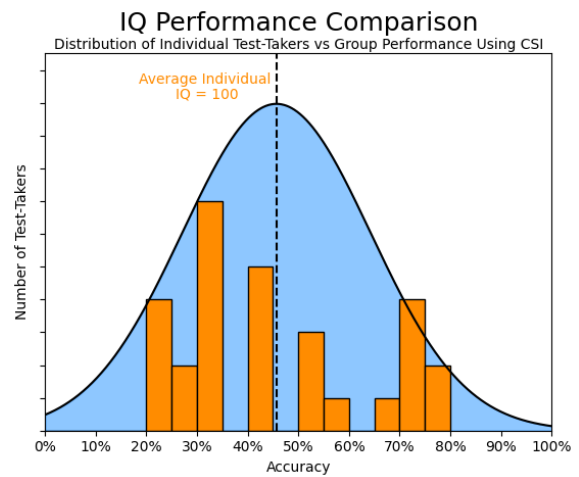

**Fig. 3.** Baseline Survey of IQ test-takers

Next, the CSI sessions were scored, and they achieved an 80.5% accuracy, corresponding to a score 1.87 standard deviations above the mean individual. Using the IQ formula above, this score corresponds to a projected collective IQ of 128. In other words, when this networked human group worked together as real-time conversational swarm, they performed 28 points higher on the IQ test than the average individual in the sample population. These results are shown below in Figure 4, compared against individuals. As shown, the CSI system scored higher in IQ, on average, than all of the individual participants tested through baseline survey.

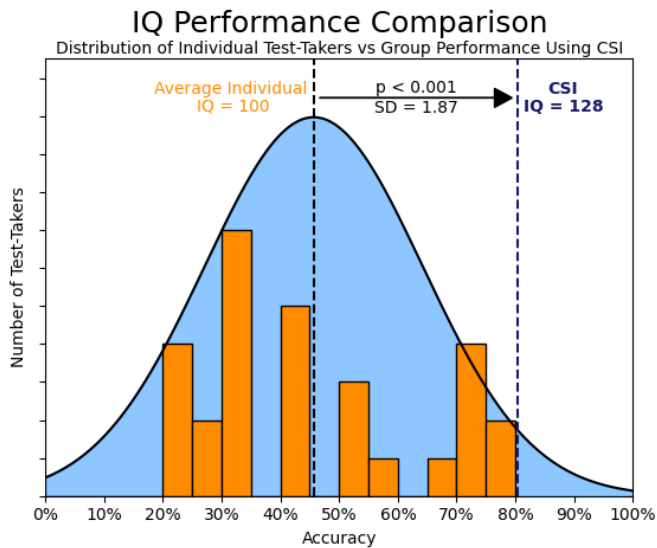

**Fig. 4.** Groups using CSI platform significantly amplify IQ

Looking next at performance versus question difficulty, we can plot how the average individual performed on easy vs hard IQ questions (orange dots in Figure 5 below) versus how the groups using CSI performed on easy vs hard questions (blue dots in the same figure). This reveals that the advantage offered by CSI technology increases with question difficulty. In fact, if we look only at the hardest 50% of questions (numbers 19 to 36), we see the average individual got 29.5% correct, while the groups using CSI averaged 70.1% correct, a 2X increase.

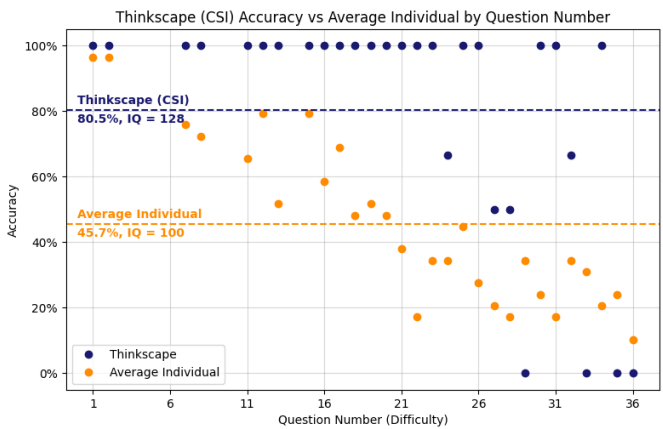

**Fig. 5.** Performance versus question difficulty for individual test-takers and groups using CSI-powered Thinkscape platform.

Turning next to statistical significance, an analysis was performed to compare the Average Individual and the real-time Conversational Swarm Intelligence (CSI) group. As shown in Table 1 below, a paired t-test was used to determine whether the increase in accuracy between the CSI groups and the Average Individual was statistically significant. The p-value was less than 0.001, showing strong evidence that on a question-by-question basis, CSI amplifies collective intelligence, enabling significantly higher accuracy than the average participant.

| Response Method | Percent Correct | % Increase in IQ over Average Individual | p Value |
| --- | --- | --- | --- |
| Average Individual | 45.7% | -- | -- |
| Thinkscape (CSI) | 80.5% | 28% | p<0.001 |

**Table 1.** IQ test scores comparing Average Individual to CSI groups.

**Assessing the Impact of the AI Agents**

As described above, the CSI-based Thinkscape platform has two distinct features compared to traditional communication platforms. First it automatically divides the sample population into set of small parallel groups called ThinkTanks™ that are optimally sized for thoughtful online conversation (4 to 7 people). Second, it adds an LLM-powered agent (called a Thinkbot™) into each of the parallel groups; each agent tasked with observing, assessing, and sharing (with other groups) conversational content based on the strength of measured confidence and conviction for that content within each local group. The experimental question is whether the increase in IQ a result of (a) breaking the population into small subgroups and aggregating sentiments locally and then globally and/or (b) intelligent information propagation across the subgroup network (using AI agents) to enable a unified conversational system that can converges on a global solution.

To assess this, the baseline IQ test data collected from isolated individuals was analyzed as follows. Using a bootstrap method, individual IQ tests were selected at random from the pool of baseline tests and organized into six subgroups of 5 or 6 individuals (with replacement). The most popular answer in each subgroup was chosen as the answer for that subgroup. The most popular answer across subgroups was chosen as the answer for the population. This method was repeated 10,000 times using the bootstrapping method, each with random selection and replacement. This gave us a statistical simulation of aggregating the raw sentiments of an example population using the unique structure of Thinktanks, but without the benefits of assessing the strength of conviction of individual members using AI agents or the benefits of intelligently propagating segments across the full network of individuals to create a unified conversation.

As shown in Figures 5, on average, the simulated subgroups, when assessed locally and aggregated globally were 64.1% accurate. Because this is a purely statistical aggregation of tests collected in isolation, we refer to this Groupwise Statistical Aggregation as a traditional Wisdom of Crowd (WoC) methodology. As expected WoC does amplify intelligence, in this case yielding an effective IQ of 115. That said, this statistical aggregation was significantly lower than the CSI methodology which yielded 80.5% accuracy on the IQ tests and achieved an effective IQ=128 (p=0.008). In other words, when using CSI the results were 26% more accurate as compared to the statistical aggregation, resulting in a 13 point increase in IQ. This suggests that Conversational Swarm Intelligence offers significant intelligence benefits, not just over the Average Individual, but over a statistical WoC method.

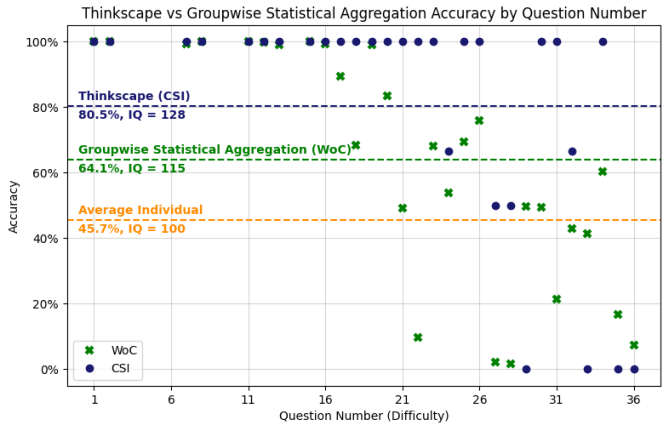

**Fig. 6.** Average Individual vs WoC vs CSI by Question Difficulty

We can also compare performance of the Average Individual, the Wisdom of Crowd (WoC) and the Conversational Swarm Intelligence (CSI) methods on the normal distribution curve expected for RAPM IQ test takers. As shown in Figure 6 below, the average individual scored in the 50th percentile (100 IQ), the bootstrapped statistical aggregation across 35 random test takers scored in the 84th percentile (115 IQ), and the groups working as a real-time conversational swarm averaged their scores in the 97th percentile (128 IQ). Furthermore, not a single individual test taker in the baseline survey scored an individual IQ as high as the average group using the Thinkscape CSI platform.

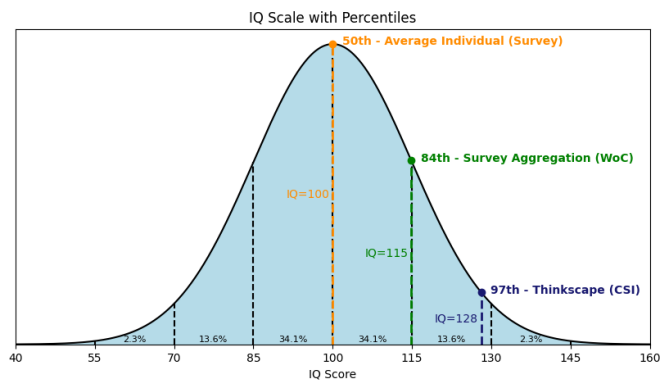

**Fig. 7.** Average Individual vs WoC vs CSI Accuracy by IQ Percentile

In addition, it is useful to compare the results of this CSI study to a previous study that tested the IQ amplification using a prior generation of Swarm AI technology that was graphical rather than conversational. Using a graphical swarming method, a 2019 study tested networked human groups using an RAPM IQ test. That study showed a 14-point increase in IQ when groups worked together as a real-time graphical swarm [19]. The current study doubled that point increase to 28 with groups working as a conversational swarm as compared to a graphical swarm. This suggests that CSI is a valuable advance in the field and has intelligence amplification benefits over prior methods.

**CSI Provides Additional Insights and Rationales**

In addition to amplifying collective intelligence, the CSI method offers additional value compared to traditional methods. That is because CSI captures a full conversational record of the deliberations along with numerical assessments of individual, groupwise, and global measures of conviction. This dataset can be analyzed to provide qualitive and quantitative insights into how and why the participants collectively converged on the solutions they did. For example, Figure 8 shows the real-time sentiment data from one question as answered by one group in support of each of the eight different answers (A through H). As shown, an incorrect answer (D, in red) was initially supported most across the network of subgroups. It was not until 62 seconds of conversation that the correct answer (G, in pink) emerged as the frontrunner, but fell back down as the group debated. It was not until 106 seconds had passed that the correct answer pulled away as the preferred solution.

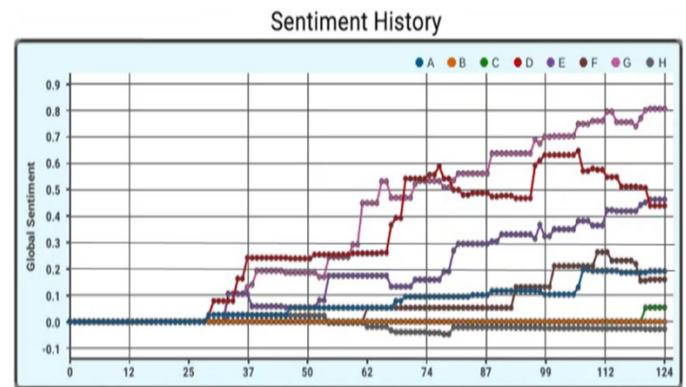

**Fig. 8.** Real-time Plot of Answer Sentiment vs Time

To better understand how the correct answer emerged across the CSI network, we can plot how insights were propagated by AI agents. In Figure 9 below, the real-time conviction within each of the 8 parallel subgroups is shown with respect to Answer H (the correct answer for that question). If a subgroup organically mentioned Answer H as a possible answer to the IQ question, a yellow light bulb is shown when a participant first argues for that solution in that subgroup. Circles and arrows depict the messages sent by the AI Agents between subgroups regarding argumentsd in favor of Answer H. When a message arrow is yellow, it represents a message introducing Answer H into that subgroup before any members had yet argued in favor of Answer H. Green arrows are shown when Answer H had already been previously supported by at least one subgroup member.

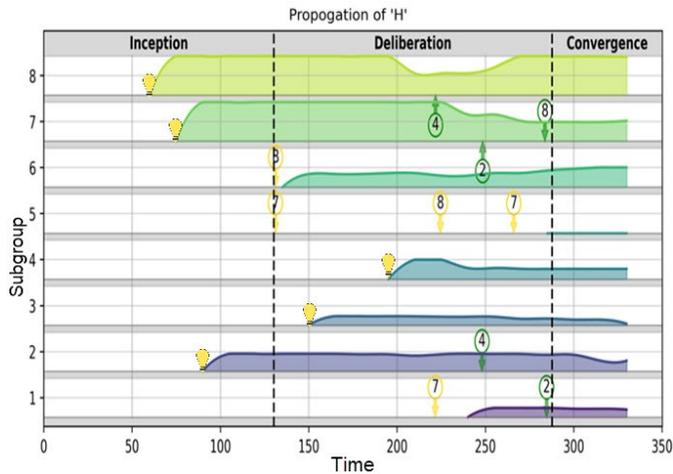

**Fig. 9**. Real-time Insight Propogation Chart across Subgroups

For clarity, Figure 9 only shows insight passing across the eight subgroups with respect to Answer H. Similar propogation charts can be generated to show the passing of insights related to each of the other answer options, whether that option was supported or disputed during the real-time deliberations. In this example, Answer H emerged over the four minute period as the option with the strongest total conviction across the CSI network. It was therefore selected by the CSI platform as the "final answer" that maximized collective confidence within the conversational swarm. The CSI platform then reports this selection and outputs the collective rationale that was converged upon during the 4-minute deliberation. In this example, the rationale output by the CSI platform was as follows.

**Rationale**: *The conversational swarm favored Answer H because the top and second rows move the fan shape counter clockwise, and when the dots and rainbow are in the same spot, it changes to blank in the bottom row. Also the first and third columns have the same pattern in the right segment, and the top right area of the circles are the same in the left and right columns. Also, it was pointed out that the top left pattern is just moving right and covering up a new section each time, and the bottom image is whatever is in the top left of the first image.*

## V. CONCLUSIONS

The results of this IQ study are promising, demonstrating that networked groups of approximately 35 individuals (a size that normally struggles to deliberate conversationally in real-time) was able to efficiently consider, debate, and converge upon answers to IQ test questions as a unified "conversational swarm" using the novel CSI structure. In addition, the results of this study show a significant amplification in collective intelligence as compared to more traditional methods. Specifically, the groups of randomly selected participants using CSI averaged a collective of score 128 on the IQ test when working together as conversational swarm intelligence, significantly outperforming both the average individual (IQ 100, $p<0.001$) and a groupwise statistical aggregation of individual tests (IQ 115, $p<0.01$).

Furthermore, the score of 128 IQ achieved by the average CSI group placed its performance in the 97[th] percentile of individual IQ test takers. In other words, only 3 of every 100 individuals taking an RAPM IQ test are likely to tie or outperform the CSI groups. In fact, none of the 35 baseline participants who took the IQ test performed as well as the CSI group. This suggests that CSI technology may be a viable pathway to achieving Collective Superintelligence, especially when expanding to larger groups in the hundreds or thousands of participants and addressing more complex and nuanced problems than standardized IQ tests.

Future research into Conversational Swarm Intelligence aims to evaluate real-time networked groups at significantly larger sizes and will test unstructured and open-ended questions that require participants to brainstorm possible solutions before deliberating and converging on preferred answers. In addition, specific use-cases such as enterprise collaboration, deliberative civic engagement, strategic priority-setting for an institution and market insights are currently being tested with CSI systems. The authors welcome collaborations with other innovators to advance research into CSI technology.